\documentclass[aps,reprint,amsmath,amssymb]{revtex4-1}
\usepackage{graphicx}
\usepackage{dcolumn}
\usepackage{bm}

\begin{document}
\title{Spin-dependent coherent transport in a double quantum dot system}
\author{L. S. Petrosyan$^{1,2}$ and T. V. Shahbazyan$^1$} 
\affiliation{
$^1$Department of Physics, Jackson State University, Jackson, Mississippi
39217 USA\\
$^2$Institute for Mathematics and High Technology, Russian-Armenian State University, 123 Hovsep Emin Street, Yerevan, 0051, Armenia
}
\begin{abstract}
We study spin-resolved resonant tunneling in a system of two quantum dots sandwiched between doped quantum wells. In the coherent (Dicke) regime, i.e., when quantum dot separation is smaller than the Fermi wavelength in a two-dimensional electron gas in quantum wells, application of an in-plane magnetic field leads to a pronounced spin-resolved structure of  conductance peak lineshape even for very small Zeeman splitting of the quantum dots' resonant levels. In the presence of electron-gas spin-orbit coupling, this spin-resolved structure is washed out due to Fermi surface deformation in the momentum space. We also show that Aharonov-Bohm flux penetrating the area enclosed by tunneling electron pathways completely destroys the conductance spin structure.
\end{abstract}
\maketitle

\section{Introduction}

Interference effects in electron transmission through localized states in semiconductor nanostructures such as, e.g., semiconductor quantum dots (QD), are among the highlights in coherent  transport studies \cite{yacoby-prl95,shuster-nature97}. The electron phase acquired in the course of tunneling through several pathways provided by QDs situated between doped semiconductor leads can result in striking features of the conductance lineshape near the transmission resonance \cite{buks-nature98}. The simplest realization of coherent transport is served by \textit{two} QDs independently coupled to a two-dimensional electron gas (2DEG) in the left and right leads while the direct tunneling between QDs is weak \cite{shahbazyan-prb94}. In a magnetic field, the conductance of such a double QD system exhibits Aharonov-Bohm oscillations \cite{blick-prl01,ensslin-prl06,hatano-prl11} as a function of magnetic flux penetrating the area enclosed by tunneling pathways \cite{shahbazyan-prb94,loss-prl00,gefen-prl01}. At zero field, the coherence between QDs is controlled by their coupling via the continuum of electronic states in the leads \cite{shahbazyan-prb94,brandes-pr05}. If QD separation $a$ is comparable to the electron Fermi wavelength in 2DEG  $\lambda_{F}$  then the electron transmission is mediated by the system eigenstates rather than by individual QDs, leading to conductance peak narrowing or Fano-like lineshapes \cite{shahbazyan-prb94,kubala-prb02,brandes-prb03,orellana-prb03}. A revealing optical analogy is cooperative emission of two excited atoms at a distance smaller than the radiation wavelength from each other (Dicke superradiance) \cite{dicke-pr54,devoe-prl96}; QD  coupling via continuum of electronic states is similar to coupling of two emitters via electromagnetic field \cite{shahbazyan-prb94,brandes-pr05,shahbazyan-prb98}.
\begin{figure}[b]
\begin{center}
\includegraphics[width=0.5\columnwidth]{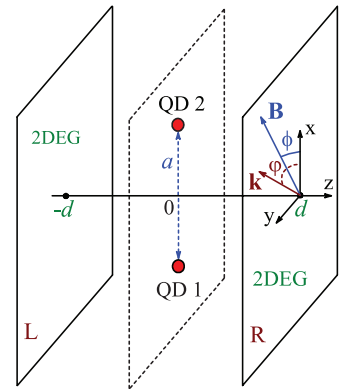}
\caption{\label{fig:1} (Color online)
Schematics of  resonant tunneling of an electron through a pair of QDs sandwiched between doped semiconductor layers in in-plane magnetic field.}
\end{center}
\end{figure}

On the other hand, spin-dependent tunneling in semiconductor nanostructures has recently attracted much interest due to the possibility of controlling simultaneously spin and charge currents in electronic circuits \cite{folk-science03}. Well-resolved spin-polarized currents were observed through single-electron or few-electron QDs subjected to an in-plane magnetic field \cite{,hanson-prl03,hanson-prb04,hanson-prl05,amasha-prb08,otsuka-prb09,stano-prb10,fujisawa-prb14}. At zero field, spin-dependent transport can be realized in semiconductor structures characterized by a strong spin-orbit (SO) coupling due to either bulk inversion asymmetry (Dresselhaus term in the Hamiltonian) or structural inversion asymmetry in the growth direction (Rashba term) \cite{datta-prl02}. In single- or double-barrier quantum well structures, the SO-induced Fermi surface splitting leads to distinct transmission coefficients for electrons with opposite spins  (spin filtering) \cite{raichev-prb03,botha-prb03,perel-prb03,tarasenko-prl04,yu-jap05,glazov-prb05,li-prb06,raikh-prb06,ye-prb07,rozhansky-prb08,smoliner-prb12}. Resonant tunneling through QDs with SO-split energy levels revealed additional structure in the conduction lineshape corresponding to spin-polarized currents \cite{apalkov-jpcm08,yokoyama-prb12}.

In this paper we study spin-dependent resonant tunneling through a double QD system sandwiched between doped semiconductor quantum wells (see Fig.\ \ref{fig:1}). Specifically, we focus on spin-resolved resonant tunneling in the Dicke regime, i.e., $ak_{F} < 1$, where  $k_{F}$ is the electron Fermi wave vector in a 2DEG. In this regime, the zero-field conductance lineshape represents a narrow peak of width $\sim (ak_{F})^{2}\Gamma$ on top of a wide peak of width $\sim 2\Gamma$, where $\Gamma$ is the single QD conduction peak width \cite{shahbazyan-prb94}. We demonstrate that an in-plane magnetic field, which introduces disbalance between spin-polarized electrons in a 2DEG, leads to a pronounced spin structure of the narrow conduction peak even for very weak Zeeman splitting of QD energy levels, $\Delta E_{0}^{z}\ll \Gamma$, i.e., when no spin splitting would normally be observed  in single QD tunneling. The   spin-resolved conductance lineshape in the Dicke regime is shown to be very sensitive to other system parameters as well, e.g., to the energy-level difference due to QD size variation. We show that SO coupling in a 2DEG, by deforming the 2DEG Fermi surfaces, \textit{suppresses} the conductance sensitivity to electron spin polarization in the Dicke regime. We also show that Aharonov-Bohm flux through the area enclosed by electron tunneling pathways completely destroys the fine spin structure of the conductance.

This paper is organized as follows. In Sec. \ref{sec:2}, we derive the general expression for the conductance in the presence of in-plane magnetic field and 2DEG SO coupling within the tunneling Hamiltonian approach. In Sec. \ref{sec:3}, we describe our analytical results for the cases when only magnetic field or only SO coupling is present.  In Sec. \ref{sec:4}, we present the results of our numerical calculation, and Sec. \ref{sec:5} concludes the paper. 

\section{Spin-dependent two-channel resonant tunneling}
\label{sec:2}

We consider electron  resonant tunneling between left and right 2DEGs located in $z=\mp d$ planes, respectively, through a pair of QDs placed in the  $z=0$ plane at a distance $a$ from each other (see Fig. \ref{fig:1}). The system is subjected to an \textit{in-plane} magnetic field ${\bf B}=B(\cos\phi,\sin\phi,0)$ characterized by vector potentials ${\bf A}^{L,R}=Ba\left(\pm \sin \phi, \mp \cos \phi, 0 \right) $ in the left/right 2DEGs, where $\phi$ is azimuthal  angle. Within the tunneling Hamiltonian approach, the system Hamiltonian is $H=H_{L}+H_{R}+H_{QD}+H_{T}$, where $H_{\alpha}$ [with $\alpha=(L,R)$] is the 2DEG Hamiltonian in the left/right  plane, $H_{QD}$ is the Hamiltonian of localized states in QD, and $H_{T}$ describes the tunneling between them. The 2DEG Hamiltonian  has the form 
\begin{align}
H_{\alpha}=\frac{1}{2m}\left(\mathbf{k}+\frac{e}{c} \mathbf{A}^{\alpha} \right) ^{2} + \beta \bm{\sigma}\times \mathbf{k} + \frac{1}{2} g\mu \left(\bm{\sigma} \cdot \mathbf{B} \right),
\end{align}
where the first, second and third terms describe, respectively, the orbital, SO, and Zeeman contributions. Here $e$, $m$, and $g$ are, respectively, electron charge, effective mass, and $g$ factor;  $c$ and $\mu_{B}$ are the speed of light and Bohr magneton; $\beta$ is the Rashba SO constant, and ${\bm \sigma}$ is the Pauli matrices vector. We assume identical left and right 2DEGs that are characterized by the same $m$ and $g$  and set $\hbar=1$ throughout. In a standard manner, by eliminating ${\bf A}^{\alpha}$ in the orbital term via gauge transformation, the 2DEG energy spectrum ${\cal E}_{\textbf{k}\delta} $ and the eigenstates $\psi_{\textbf{k}\delta}^{\alpha}(\textbf{r})$ in each plane can be found as
\begin{equation}
\label{eigen}
{\cal E}_{\textbf{k}\delta} =\frac{k^{2}}{2m}+\delta |\xi_{ \textbf{k}}|,
~
\psi_{\textbf{k}\delta}^{L,R}(\textbf{r})=\frac{e^{i{\bf k}{\bf r}\pm i\frac{e}{c}{\bf B}\times {\bf r}}}{\sqrt{2A}}
\left(\! 
\begin{array}{c}
1\\
\delta e^{i\theta_{\textbf{k}}}
\end{array}
\!\right) ,
\end{equation}
where  the variable
\begin{equation}
\label{xi}
\xi_{\textbf{k}}\equiv |\xi_{\textbf{k}}|e^{i\theta_{\textbf{k}}}=e^{i\phi}\omega_{z}/2 -i\beta k e^{i\varphi}
\end{equation}
depends both on the orientation of the wave vector, $\varphi=\arg(\mathbf{k})$, and on the magnetic field orientation $\phi$ relative to the $x$ axis. Here $\omega_{z}=g\mu_{B} B$ is 2DEG Zeeman energy, $\delta=\pm 1$ is chirality,  and $A$ is the normalization area. Two possible signs ($\pm$) of magnetic phase in Eq.~(\ref{eigen}) correspond to the left/right 2DEG, respectively. For each chirality $\delta=\pm 1$, the Fermi surface, $k_{F}^{\delta}(\varphi)$ represents closed contours in ${\bf k}$ space satisfying 
\begin{equation}
\label{fermi-surface}
\frac{k^{2}}{2m}\pm |\xi_{\bf k}(\phi)|=E_{F},
\end{equation}
where $E_{F}$ is the Fermi energy. In the presence of SO coupling, the Fermi-surface shape  depends on the magnetic field orientation $\phi$. 

The tunneling Hamiltonian describing transitions between QD and 2DEG states has the form
\begin{align}
H=\sum_{jss'} E_{j}^{ss'}c_{js}^{\dagger}c_{js'}+ 
\sum_{\textbf{k}\delta\alpha}
{\cal E}_{\textbf{k}\delta}^{\alpha}c_{\textbf{k}\delta\alpha}^{\dagger}c_{\textbf{k}\delta\alpha}
~~~~~~~~~~~~~~~
\nonumber\\
+\sum_{\textbf{k}\delta \alpha js} \left (V_{\textbf{k}\delta\alpha}^{js} c_{js}^{\dagger}c_{\textbf{k}\delta\alpha}+\text{H.c.}\right ),
\end{align}
where $E_{j}^{ss'}=E_{j}\delta_{ss'}+\frac{1}{2}g_{0}\mu_{B}\left ({\bf B}\cdot {\bm \sigma}\right )_{ss'}$ is the QD energy matrix. Here  $E_{j}$ ($j=1,2$) are QD resonant energy levels,  $g_{0}$ is the QD $g$ factor (the spin quantization axis is chosen along $\textbf{z}$ and $s=\pm 1$ corresponds to spin up/down projections), and $V_{\textbf{k}\delta\alpha}^{js}$ is the electron transition matrix element between QD state $|js\rangle$ and 2DEG state $|\textbf{k}\delta\alpha\rangle$ ($\alpha=L,R)$. We assume no spin flip during tunneling in the  lateral direction. 

Within the tunneling Hamiltonian approach, the conductance is given by \cite{brandes-pr05}
\begin{equation}
\label{cond}
G=\frac{e^{2}}{\pi\hbar}\text{Tr}\left(\hat{\Gamma}^{R}\frac{1}{E_{F}-\hat{E}-\hat{\Sigma}} \,\hat{\Gamma}^{L}\frac{1}{E_{F}-\hat{E}-\hat{\Sigma}^{\dagger}}\right),
\end{equation}
where  the matrix $E_{jk}^{ss'}=\delta_{jk} E_{j}^{ss'}$ is diagonal in QD indices, $\hat{\Sigma}=\hat{\Sigma}_{L}+\hat{\Sigma}_{R}$ is the QD self-energy matrix due to the transitions to the left and right 2DEG,
\begin{equation}
\label{sigma0}
\left (\Sigma_{\alpha}\right )_{jk}^{ss'}\equiv \left (\Delta_{\alpha}\right )_{jk}^{ss'}-\frac{i}{2} \left (\Gamma_{\alpha}\right )_{jk}^{ss'}=   \sum_{{\bf k}\delta}\dfrac{V_{\textbf{k}\delta\alpha}^{js}V_{\textbf{k}\delta\alpha}^{ks'*}}{E_{F}-{\cal E}_{{\bf k}\delta}+i0},
\end{equation}
 and the trace is taken in both configuration and spin space. The transition matrix element can be presented as \cite{shahbazyan-prb94} $V_{\textbf{k}\delta\alpha}^{js}=t_{\alpha}\psi_{\textbf{k}\delta}^{\alpha s}(\textbf{r}_{j})$, where 
${\bf r}_{j}$ are the \textit{in-plane} projection of QD coordinates and $t_{\alpha}$ is the tunneling amplitude between QD and 2DEG (we assume that the potential barrier is sufficiently high and neglect the $t_{\alpha}$ dependence on energy). The self-energy (\ref{sigma0}) then takes the form 
\begin{equation}
\label{sigma-green}
\left (\Sigma_{\alpha}\right )_{ij}^{ss'} =t_{\alpha}^{2}G_{\alpha}^{ss'}({\bf r}_{i}-{\bf r}_{j}),
\end{equation}
where
\begin{align}
\label{green}
G^{ss'}_{L,R}(\textbf{r}_{i}-\textbf{r}_{j})=\frac{1}{2}e^{ \pm i(1-\delta_{ij})(ad/l^2)\sin\phi} \sum_{\delta=\pm}\delta^{(s-s')/2}
\nonumber\\
\times\int \frac{d^2 k}{(2\pi )^2 }  \frac{e^{i \mathbf{k}(\mathbf{r}_i-\mathbf{r}_j)+i\theta_{\bf k} (s'-s)/2}} {E_F-{\cal E}_{\textbf{k}\delta}+i0} 
\end{align}
is the 2DEG Green's function corresponding to eigenstates (\ref{eigen}), $l=\sqrt{c/eB}$ is the magnetic length, and $\theta_{\bf k}=\arg(\xi_{\bf k})$. The decay matrix $\hat{\Gamma}$ and energy shift matrix $\hat{\Delta}$,  which are determined, respectively, by the singular and principal parts of  the Green's function (\ref{green}), represent $4\times 4$ matrices in spin and configuration space. Note that the electron Green's function (\ref{green})  is known explicitly in the presence of  either magnetic field or SO coupling  but not both.

\section{Calculation of conductance}
\label{sec:3}
We consider a symmetric case when two QDs with  resonant levels $E_{1}=E_{2}=E_{0}$ at a distance $a$ from each are separated by a tunneling barrier of thickness $d$ from identical 2DEGs (see Fig. \ref{fig:1}). In this case, the tunneling amplitudes are equal, $t_{L}=t_{R}=t$, and the QD self-energies due to tunneling to left/right 2DEG differ only by the Aharonov-Bohm phase factor: $\left (\Sigma_{L,R}\right )_{ij}^{ss'}=e^{ \pm i(1-\delta_{ij})(ad/l^2)\sin\phi}\tilde{\Sigma}_{ij}^{ss'}$, with
\begin{align}
\label{sigma}
\tilde{\Sigma}_{ij}^{ss'}=\frac{t^{2}}{2} \sum_{\delta=\pm 1}\delta^{\frac{s-s'}{2}}\int \frac{d^2 k}{(2\pi )^2 }  \frac{e^{i \mathbf{k}\cdot\mathbf{r}_{ij}+i\theta_{\bf k} (s'-s)/2}} {E_F-k^2/2m-\delta |\xi_{\bf k}|+i0},
\end{align}
where $\xi_{\textbf{k}}\equiv |\xi_{\textbf{k}}|e^{i\theta_{\textbf{k}}}$ is given by Eq.\ (\ref{xi}) and we denoted $\mathbf{r}_{ij}=\mathbf{r}_i-\mathbf{r}_j$.

\subsection{Dicke tunneling in the presence of in-plane magnetic field}
\label{sec:3a}

Consider first the conductance in the absence of SO coupling in 2DEGs ($\beta=0$). The Fermi surface, as  determined by  Eq. (\ref{fermi-surface})  with $\xi_{\textbf{k}}=e^{i\phi}\omega_{z}/2$, represents two circles in momentum space with radii $k_{F}^{\delta}$ given by
\begin{align}
\label{fermi-surface1}
k_{F}^{\pm}=k_{F}\sqrt{1\mp\omega_{z}/2E_{F}},
\end{align}
where $k_{F}=\sqrt{2mE_{F}}$.  In this case we have $\theta_{\textbf{k}}=\phi$, and so the  spin and orbital degrees of freedom in the self-energy (\ref{sigma}) factorize,
\begin{align}
\label{sigma1}
\tilde{\Sigma}_{ij}^{ss'}=-\frac{i\Gamma}{4} \sum_{\delta=\pm 1}\delta^{\frac{s-s'}{2}} e^{i\phi(s-s')/2} H_{0}^{(1)}\left (k_{F}^{\delta}r_{ij}\right ),
\end{align}
where $H_{0}^{(1)}(x)$ is the Hankel function of  the first kind and $\Gamma=mt^{2}$ is the resonant level spectral width for an isolated QD due to tunneling to the 2DEG (see below). The function $H_{0}^{(1)}\left (k_{F}^{\delta}r_{ij}\right )$ can be viewed as a $2\times 2$ matrix in configuration space with diagonal elements $H_{0}^{(1)}\left (0\right )$ and nondiagonal elements $H_{0}^{(1)}\left (k_{F}^{\delta}a\right )$. Note that $\text{Im} \left[H_{0}^{(1)} (0)\right ]$ contains logarithmic divergence that should be properly regularized. Namely, for infinitesimal $r_{ij}=\epsilon\rightarrow 0$, we have 
$H_{0}^{(1)}\left (k_{F}^{\delta}\epsilon\right )\approx 1+(2i/\pi)\left [\gamma_{E}+ \ln \left (k_{F}^{\delta}\epsilon/2\right )\right ]$, where $\gamma_{E}$ is the Euler constant. We now subtract the \textit{zero-field} value of $\text{Im} \left [ H_{0}^{(1)} (k_{F}^{\delta}\epsilon)\right ]$, i.e., with $k_{F}^{\delta}=k_{F}$, so that the regularized expression for $H_{0}^{(1)} (0)$ is $1+(2i/\pi)\ln \left (k_{F}^{\delta}/k_{F} \right )$. Such regularization corresponds to zero energy shift for an \textit{isolated} QD in the absence of magnetic field;  hereafter, we will  use only regularized quantities. The matrix $H_{0}^{(1)}\left (k_{F}^{\delta}r_{ij}\right )$ can be written in terms of the Pauli matrices in configuration space $\bm{\tau}$ as
\begin{align}
H_{0}^{(1)}\left (k_{F}^{\delta}r_{ij}\right )=&\left [1+\frac{2i}{\pi}\ln \left (\frac{k_{F}^{\delta}}{k_{F}} \right )\right ]I_{\tau}
\nonumber\\
&+\left [J_{0}\left (k_{F}^{\delta}a\right )+iY_{0}\left (k_{F}^{\delta}a\right )\right ]\tau_{1},
\end{align}
where $I_{\tau}$ is the unit matrix  and we used $H_{n}^{(1)}(x)=J_{n}(x)+iY_{n}(x)$, with $J_{n}$ and $Y_{n}$ being Bessel functions of the first and second kinds, respectively. Expressing the spin factor in Eq. (\ref{sigma1})  via Pauli spin matrices as 
\begin{equation}
S_{\delta}^{ss'}(\phi)\equiv \delta^{\frac{s-s'}{2}} e^{i\phi(s-s')/2}=\left (I_{\sigma}+\delta {\bm \sigma}\cdot \hat{\bm b}\right)_{ss'},
\end{equation}
 where $\hat{\bm b}$ is a unit vector along the magnetic field ($I_{\sigma}$ is the unit matrix in spin space), the self-energy $\hat{\Sigma}_{\alpha}=\hat{\Delta}_{\alpha}-i\hat{\Gamma}_{\alpha}/2$ is presented as a $4\times 4$ matrix in spin and configuration space with 
\begin{align}
\label{gamdel}
&\hat{\Gamma}_{L,R}=\frac{\Gamma}{2} \sum_{\delta=\pm 1}
S_{\delta}\otimes \bigl [I_{\tau}+ \Phi_{L,R}J_{0} (k_{F}^{\delta}a )\tau_{1} \bigr ],
\\
&\hat{\Delta}_{L,R}=\frac{\Gamma}{4} \sum_{\delta=\pm 1}
S_{\delta}\otimes \Bigl [ \frac{2}{\pi}\ln \left (\frac{k_{F}^{\delta}}{k_{F}} \right ) I_{\tau}+ \Phi_{L,R}Y_{0} (k_{F}^{\delta}a )\tau_{1} \Bigr ],
\nonumber
\end{align}
where $\Phi_{L,R}=e^{ \pm i(ad/l^2)\sin\phi}$ is the Aharonov-Bohm factor. Using these expressions, the conductance (\ref{cond}) can be straightforwardly evaluated.

Since orbital and spin sectors in Eq.\ (\ref{gamdel}) factorize, they can be diagonalized independently, and an explicit expression for the conductance can be obtained. For simplicity, consider magnetic field directed along the $x$ axis (i.e., $\phi=0$). In this case, there is no Aharonov-Bohm flux ($\Phi_{L,R}=1$), so that  $\Sigma_{L}=\Sigma_{R}$, and after simple algebra, we obtain
\begin{equation}
\label{cond1}
G=\frac{e^{2}}{\pi\hbar}\sum_{q,p=\pm}\frac{\Gamma_{q p}^{2}}{\left (E_{F}-E_{0}-p\,\Delta E_{0}^{z}/2-\Delta_{q p}\right )^{2}+\Gamma_{q p}^{2}},
\end{equation}
where $\Delta E_{0}^{z}$ is the QD Zeeman energy and
\begin{align}
\label{gamdel1}
&\Gamma_{qp}=\Gamma \bigl [1+ qJ_{0} (k_{F}^{p}a )\bigr ],
\\
&\Delta_{q p}=\Gamma\Bigl [ \frac{1}{\pi}\ln \left (1- p\frac{\omega_{z}}{2E_{F}} \right ) + qY_{0} (k_{F}^{p}a ) \Bigr ],
\nonumber
\end{align}
with Fermi momenta $k_{F}^{p}$ corresponding to the two Fermi surfaces ($p=\pm$) given by Eq. (\ref{fermi-surface1}). 

For zero field, i.e., $\omega_{z}=\Delta E_{0}^{z}=0$ and $k_{F}^{p}=k_{F}$, both spin channels contribute equally, and we recover the known result for Dicke tunneling through a pair of QDs \cite{shahbazyan-prb94},
\begin{equation}
\label{cond2}
G=\frac{2e^{2}}{\pi\hbar}\sum_{\pm}\frac{\Gamma_{\pm}^{2}}{\left (E_{F}-E_{0}-\Delta_{\pm}\right )^{2}+\Gamma_{\pm}^{2}},
\end{equation}
where $\Gamma_{\pm}=\Gamma\left [1\pm J_{0} (k_{F}a )\right ]$ and $\Delta_{\pm}=\pm \Gamma Y_{0} \left (k_{F}a\right )$. For $k_{F}a\ll 1$, the conductance lineshape represents a narrow peak of width $\Gamma_{-}\approx (k_{F}a )^{2}\Gamma/4$ on top of a wide peak of width $\Gamma_{+}\approx 2\Gamma$. With magnetic field turned on, each Zeeman-split Fermi surface described by Eq. (\ref{fermi-surface1}) contributes independently to the conductance (\ref{cond1}), thereby giving rise to the spin fine structure of peak conductance. With a further field increase, the Fermi surface $k_{F}^{-}$ shrinks to a point, and for $\omega_{z}>2E_{F}$ the 2DEG is fully spin polarized. In the proximity of the critical field, the conductance lineshape undergoes  dramatic changes, as illustrated in the next section.

\subsection{Dicke tunneling in the presence of SO coupling in 2DEG}
\label{sec:3b}

Consider now the case of a 2DEG with Rashba SO coupling at zero magnetic field. In this case, an analytical expression for the electron Green's function is well known and Eq.\ (\ref{sigma-green})  takes the form (we suppress spin indices)
\begin{align}
\label{sigma2}
\tilde{\Sigma}_{ij}=\Sigma_{ij}^{(0)}I_{\sigma}+\Sigma_{ij}^{(1)}\left (\hat{\textbf{z}}\times \hat{\textbf{r}}_{ij}\right )\cdot {\bm \sigma},
\end{align}
where $\hat{\textbf{r}}_{ij}$ and $\hat{\textbf{z}}$ are unit vectors along $\textbf{r}_{ij}$ and the $z$ axis and $\Sigma_{ij}^{(0)}$ and $\Sigma_{ij}^{(1)}$ are matrices in configuration space,
\begin{align}
\label{sigma3}
\tilde{\Sigma}_{ij}^{(0)}=-\frac{i\Gamma}{4} \sum_{\delta=\pm 1}\frac{k_{F}^{\delta}}{\tilde{k}_{F}}\, H_{0}^{(1)}\left (k_{F}^{\delta}r_{ij}\right ),
\\
\tilde{\Sigma}_{ij}^{(1)}=-\frac{\Gamma}{4} \sum_{\delta=\pm 1}\delta\,\frac{k_{F}^{\delta}}{\tilde{k}_{F}}\, H_{1}^{(1)}\left (k_{F}^{\delta}r_{ij}\right ).
\end{align}
Here $k_{F}^{\delta}=\tilde{k}_{F}-\delta k_{R}$ (with $\delta=\pm 1$) are solutions of Eq. (\ref{fermi-surface}) describing two Fermi surfaces, $k_{R}=m\beta$ is the characteristic momentum associated with Rashba SO coupling and $\tilde{k}_{F}=\sqrt{k_{F}^{2}+k_{R}^{2}}$. Expressing the above matrices via Pauli matrices in configuration space, we obtain (after regularization)

\begin{align}
\label{sigma4}
&\tilde{\Sigma}_{ij}^{(0)}= \frac{\Gamma}{4}\sum_{\delta=\pm 1}\frac{k_{F}^{\delta}}{\tilde{k}_{F}}\left [\left (\frac{2}{\pi}\ln\frac{k_{F}^{\delta}}{k_{F}}-i\right )I_{\tau} - iH_{0}^{(1)}(k_{F}^{\delta}a)\tau_{1}\right ],
\nonumber\\
&\tilde{\Sigma}_{ij}^{(1)}=-\frac{\Gamma}{4}\sum_{\delta=\pm 1}\delta\,\frac{k_{F}^{\delta}}{\tilde{k}_{F}}\, H_{1}^{(1)}\left (k_{F}^{\delta}a\right ) \tau_{1}.
\end{align}
For QDs placed along the $x$ axis, $\hat{\textbf{r}}_{ij}=-\hat{\textbf{r}}_{ji}=\hat{\textbf{x}}$, the $4\times 4$ self-energy matrix $\hat{\Sigma}_{\alpha}=\hat{\Delta}_{\alpha}-i\hat{\Gamma}_{\alpha}/2$ can be explicitly obtained as
\begin{align}
\label{gamdel2}
&\hat{\Gamma}_{\alpha}=\Gamma \left (I_{\sigma}\otimes I_{\tau}\right ) + \Gamma_{1} \left (I_{\sigma}\otimes \tau_{1}\right ) +\Gamma_{2} \left (\sigma_{2}\otimes \tau_{2}\right ),
\nonumber\\
&\hat{\Delta}_{\alpha}=\Delta_{0}\left (I_{\sigma}\otimes I_{\tau} \right )+ \Delta_{1} \left (I_{\sigma}\otimes \tau_{1}\right ) +\Delta_{2} \left (\sigma_{2}\otimes \tau_{2}\right ),
\end{align}
where
\begin{align}
\label{gamdel3}
&\Gamma_{1}=\frac{\Gamma}{2}\sum_{\delta=\pm 1}\frac{k_{F}^{\delta}}{\tilde{k}_{F}}\,J_{0}(k_{F}^{\delta}a),
~~
\Gamma_{2}=-\frac{\Gamma}{2}\sum_{\delta=\pm 1}\delta\,\frac{k_{F}^{\delta}}{\tilde{k}_{F}}\,J_{1}(k_{F}^{\delta}a),
\nonumber\\
&\Delta_{0}= \frac{\Gamma}{4}\sum_{\delta=\pm 1}\frac{k_{F}^{\delta}}{\tilde{k}_{F}}\frac{2}{\pi}\ln\frac{k_{F}^{\delta}}{k_{F}},
~~
\Delta_{1}= \frac{\Gamma}{4}\sum_{\delta=\pm 1}\frac{k_{F}^{\delta}}{\tilde{k}_{F}}\,Y_{0}(k_{F}^{\delta}a),
\nonumber\\
&\Delta_{2}=-\frac{\Gamma}{4}\sum_{\delta=\pm 1}\delta\,\frac{k_{F}^{\delta}}{\tilde{k}_{F}}\,Y_{1}(k_{F}^{\delta}a).
\end{align}
By setting the SO coupling to zero, i.e., $k_{F}^{\delta}=\tilde{k}_{F}=k_{F}$, all quantities in Eq. (\ref{gamdel3}) have vanished, except $\Gamma_{1}$ and $\Delta_{1}$, so that the conductance (\ref{cond2}) is recovered. Note, however, that, in the presence of SO coupling, no explicit formula for the conductance can be derived and Eq. (\ref{cond}) still needs to be evaluated numerically.

In the presence of both magnetic field and SO coupling, no analytical expression for the self-energy matrix (\ref{sigma2}) is available . The results of our numerical calculations of the conductance are presented in the next section.

\section{Numerical results and discussion}
\label{sec:4}

Here we describe our results for the conductance (\ref{cond}) obtained by numerically evaluating the matrix elements (\ref{sigma}).  To simplify the analysis, we assume identical 2DEGs in the left and right planes with Zeeman energy $\omega_{z}$ and with Rashba SO coupling described by characteristic momentum $k_{R}$. In the symmetric configuration, two QDs separated by a distance $a$ are located  in the middle between 2DEG planes (see Fig. \ref{fig:1}). The resonant levels in QDs have energies $E_{0}\pm \Delta E_{0} \pm\Delta E_{0}^{z}$, where $\Delta E_{0}$ is a shift from medium level energy $E_{0}$ due to variations in QD size and $\Delta E_{0}^{z}$ is QD Zeeman splitting. Here we disregard SO splitting of QD levels \cite{apalkov-jpcm08} and instead focus on the role of 2DEG SO coupling. Rather than restricting ourselves to a specific material, we presents our numerical results for a wide range of parameters to describe comprehensively the role of magnetic field and SO  coupling in coherent transport in a double QD system.

\begin{figure}[tb]
\begin{center}
\includegraphics[width=0.9\columnwidth]{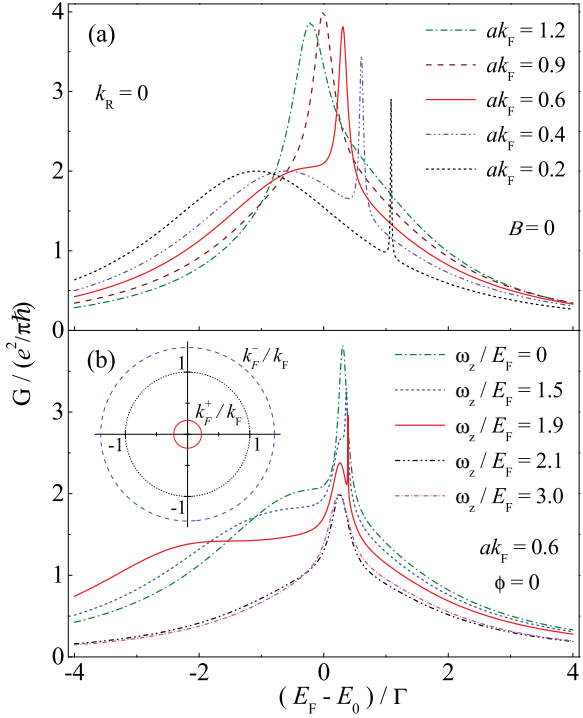}
\caption{\label{fig:2} (Color online)
Conductance through identical QDs ($\Delta E_{0}=0$) is shown for (a) several QD separations  $a$  and (b) several values of 2DEG Zeeman energy  $\omega_{z}$  in the absence of 2DEG SO coupling and QD Zeeman splitting. The inset shows spin-split 2DEG Fermi surfaces in the $\textbf{k}$ plane for $\omega_{z}/E_{F}=1.9$.
}
\end{center}
\end{figure}

We start with the case of zero SO coupling ($k_{R}=0$). In Fig. \ref{fig:2}(a) we show zero-field conductance vs. Fermi energy for several values of  electron concentration (or QD separation) as the parameter $ak_{F}$ traverses the region $ak_{F}\lesssim 1$ (since $\Gamma/E_{F}\ll 1$, the parameter $ak_{F}$ is nearly constant in the resonance region). While for $ak_{F}>1$ the conductance shows a single peak of amplitude $\sim 4$ (in units of $e^{2}/\pi\hbar$) corresponding to two orbital and two spin channels, with decreasing $ak_{F}$ it develops a double-peak structure with a narrow peak on top of a wider peak. This is a characteristic signature of coherent Dicke tunneling \cite{shahbazyan-prb94} due to electron transmission through symmetric and antisymmetric superpositions of QD states, rather than through individual QDs, with the respective rates $\Gamma_{\pm}=\Gamma \left [1\pm J_{0}(ak_{F})\right ]$. The resonance shift that takes place with decreasing $ak_{F}$ is caused by QD level repulsion [$\Delta_{\pm}=\pm \Gamma Y_{0} \left (k_{F}a\right )$] due to QD coupling through the 2DEG, while the peak narrowing is due to weaker coupling of the antisymmetric state to 2DEGs as the electron Fermi wavelength $\lambda_{F}=2\pi/k_{F}$ exceeds QD separation $a$.

\begin{figure}[tb]
\begin{center}
\includegraphics[width=0.9\columnwidth]{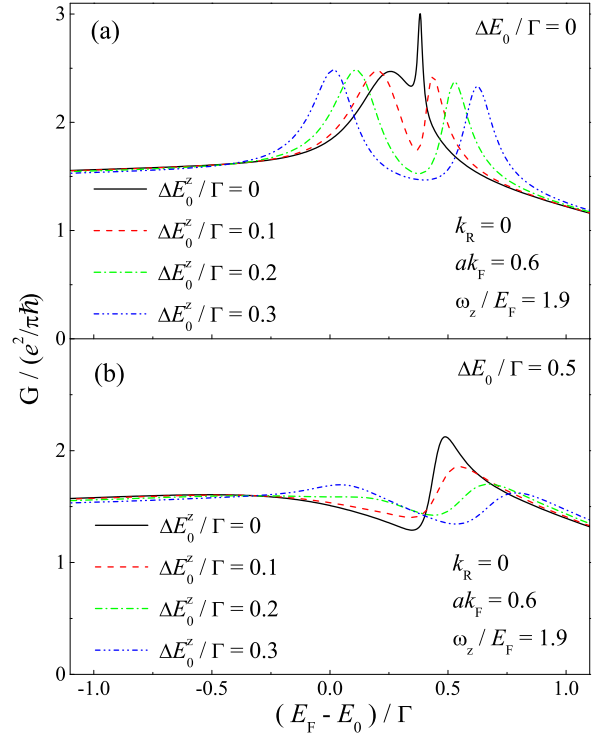}
\caption{\label{fig:3} (Color online)
Conductance peak evolution in the absence of 2DEG SO coupling is shown with increasing QD Zeeman splitting $\Delta E_{0}^{z}$ for QD level detuning values (a) $\Delta E_{0}=0$ and (b) $\Delta E_{0}=0.5\Gamma$ at the near-critical value of in-plane field $\omega_{z}=1.9E_{F}$.
}
\end{center}
\end{figure}
\begin{figure}[tb]
\begin{center}
\includegraphics[width=0.9\columnwidth]{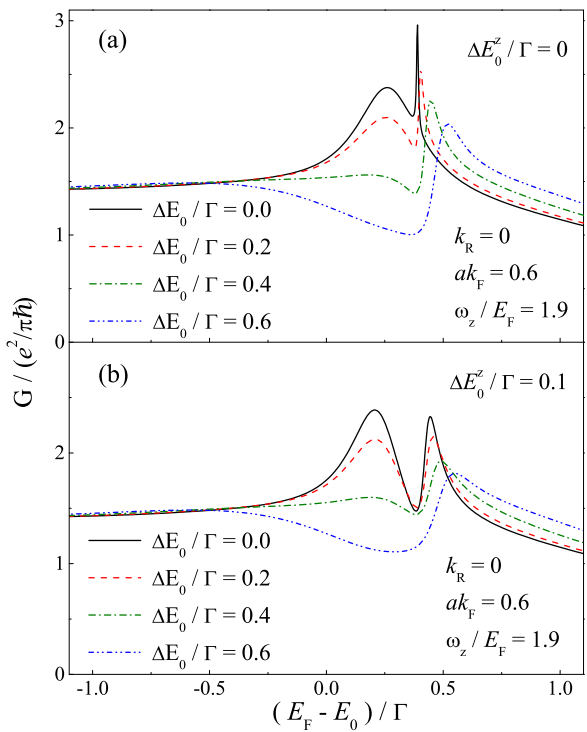}
\caption{\label{fig:4} (Color online)
Conductance peak evolution in the absence of 2DEG SO coupling is shown with increasing QD level detuning $\Delta E_{0}$ for QD Zeeman splitting values (a) $\Delta E_{0}^{z}=0$ and (b) $\Delta E_{0}^{z}=0.1\Gamma$ at the near-critical value of  in-plane field $\omega_{z}=1.9E_{F}$.
}
\end{center}
\end{figure}

In Fig.\ \ref{fig:2}(b) we show the conductance evolution in the Dicke regime (i.e., for $ak_{F}<1$) with changing in-plane magnetic field. To distinguish between various effects of magnetic field, here we chose $\textbf{B}\parallel \hat{\textbf{x}}$ (i.e., $\phi=0$ and, hence, no Aharonov-Bohm flux) and, for a moment, disregard  QD Zeeman splitting ($\Delta E_{0}^{z}=0$). With increasing 2DEG Zeeman energy $\omega_{z}$, the narrow conductance peak develops a shoulder and then  splits into two peaks of the width $\Gamma_{-}^{\pm}=\Gamma \left [1- J_{0}(ak_{F}^{\pm})\right ]$ corresponding to two spin-polarized antisymmetric states with energies shifted by $\Delta_{-}^{\pm}=- \Gamma Y_{0} \left (ak_{F}^{\pm}\right )$. This splitting is caused by tunnel coupling of QD levels to spin-polarized electrons in the 2DEG with different Fermi momenta [see inset in Fig.\ \ref{fig:2}(b)]. With increasing field, as  $\omega_{z}/2$ approaches $E_{F}$, the upper spin subband becomes nearly empty, while the lower spin subband population nearly doubles; the emergence of smaller and larger Fermi momenta, $k_{F}^{+}$ and $k_{F}^{-}$,  leads to a significant difference in the new peaks' width. A similar effect takes place for tunneling through the symmetric state; however, the spin-polarized  states with wide widths $\Gamma_{+}^{\pm}=\Gamma \left [1+ J_{0}(ak_{F}^{\pm})\right ]$ and energy shifts $\Delta_{+}^{\pm}= \Gamma Y_{0} \left (ak_{F}^{\pm}\right )$ are not well resolved and manifest themselves as extended plateaus on the  low-energy side. With further field increase, as the upper spin subband is completely depopulated ($\omega_{z}>2E_{F}$), the tunneling current is fully spin polarized,  and the conductance shows only a single peak.

In Fig. \ref{fig:3}, we show the effect of QD Zeeman splitting $\Delta E_{0}^{z}$ on  the conductance lineshape. Here we focus on QD level spin splitting \textit{per se} and therefore only change  QD $g$ factor while keeping the  magnetic field constant. To highlight coherent effects in spin-resolved tunneling, we chose very small values of QD Zeeman splitting  ($\Delta E_{0}^{z}/\Gamma\ll 1$) that normally would not be resolved in single QD tunneling, and we plot the  narrow conductance peak lineshape for nearly critical field ($\omega_{z}/E_{F}=1.9$);  the effect of small $\Delta E_{0}^{z}$ on the wide conductance peak is negligible. Remarkably, the narrow peak exhibits a pronounced spin splitting for $\Delta E_{0}^{z}$ as small as $0.1\Gamma$. 
With increasing $\Delta E_{0}^{z}$, this splitting steadily increases, with peak-to-peak separation being $\sim 2\Delta E_{0}^{z}$. At the same time, the overall lineshape  becomes more symmetrical as the QD upper spin levels now couple to higher $k_{F}^{-}$ 2DEG states [see Fig. \ref{fig:3}(a)]. Figure \ref{fig:3}(b) shows the conductance peak evolution with changing $\Delta E_{0}^{z}$ when the QD level energies are slightly different, $E_{1,2}=E_{0}\pm \Delta E_{0}$, e.g., due to QD size variation.  For  $\Delta E_{0}=0.5 \Gamma$, an overall drop in peak amplitude is observed, and with increasing $\Delta E_{0}^{z}$, the conductance exhibits no sharp features.

\begin{figure}[tb]
\begin{center}
\includegraphics[width=0.9\columnwidth]{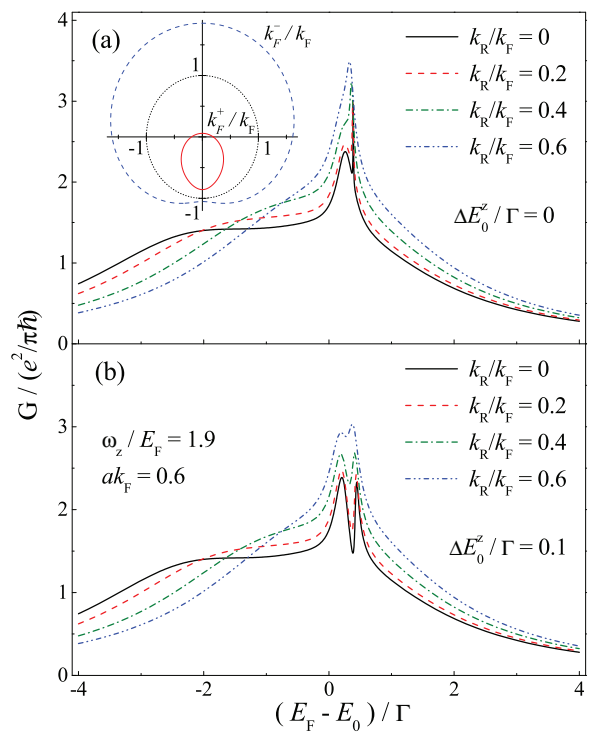}
\caption{\label{fig:5} (Color online)
Conductance evolution with increasing 2DEG SO coupling is shown for QD Zeeman splitting values (a)  $\Delta E_{0}^{z}=0$  and (b)  $\Delta E_{0}^{z}=0.1\Gamma$ at the near-critical in-plane magnetic field, $\omega_{z}=1.9E_{F}$, and $\Delta E_{0}=0$. The inset shows spin-split 2DEG Fermi surfaces in the $\textbf{k}$-plane for $\omega_{z}/E_{F}=1.9$.
}
\end{center}
\end{figure}
\begin{figure}[tb]
\begin{center}
\includegraphics[width=0.9\columnwidth]{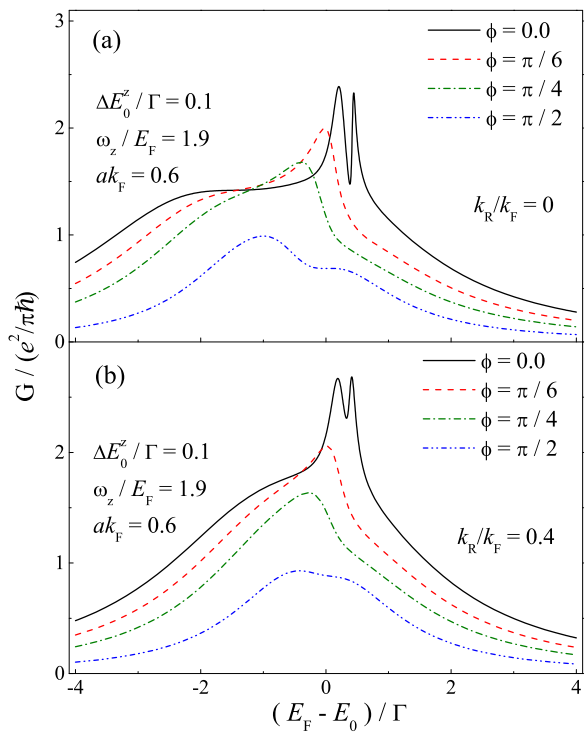}
\caption{\label{fig:6} (Color online)
Conductance evolution with increasing Aharonov-Bohm flux is shown with changing magnetic field tilt angle $\phi$ (a) in the absence and (b) in the presence  of 2DEG SO coupling at $\Delta E_{0}^{z}=0.1\Gamma$,  $\omega_{z}=1.9E_{F}$, and $\Delta E_{0}=0$.
}
\end{center}
\end{figure}

In Fig. \ref{fig:4}, we show a complementary case of conductance peak evolution with changing $\Delta E_{0}$ at a constant $\Delta E_{0}^{z}$ for nearly critical magnetic field strength ($\omega_{z}/E_{F}=1.9$). For finite $\Delta E_{0}$, the narrow peak of the upper spin subband gets wider as the upper QD level $E_{1}$ now couples to 2DEG states with a higher Fermi momentum $k_{F}^{-}$ [see Fig. \ref{fig:4}(a)]. With increasing $\Delta E_{0}$, the low-energy resonance disappears and turns into antiresonance; this effect is similar to the zero-field case \cite{shahbazyan-prb94}. A similar evolution of narrow peak lineshape is observed for a finite QD Zeeman splitting [see Fig. \ref{fig:3}(b)].

We now turn to the combined effect on Dicke tunneling of 2DEG SO coupling and in-plane magnetic field. At zero field, the tunneling matrix elements can be explicitly calculated (Sec. \ref{sec:3b}); however,  for a realistic range of parameters, the 2DEG SO coupling is relatively weak, $k_{R}=\beta m < k_{F}$, and has no significant effect on conductance lineshape. The situation changes in the presence of nearly critical in-plane magnetic field, i.e., when the 2DEG is nearly spin polarized. In Fig.\ \ref{fig:5}, we show the evolution of the conductance lineshape with increasing 2DEG SO coupling at  $\omega_{z}/E_{F}=1.9$ and $\phi=0$. For $\Delta E_{0}^{z}=0$, the spin splitting of the narrow peak disappears with increasing $k_{R}$, and for $k_{R}/k_{F}> 0.5$ the two  peaks merge [see Fig.\ \ref{fig:5}(a)]. A similar evolution takes place in the presence QD Zeeman splitting [see Fig.\ \ref{fig:5}(b)]; the pronounced dip due to combined 2DEG and QD Zeeman effects evolves into a small dent. Such a behavior can be traced to the change in Fermi-surface shape in the presence of both 2DEG SO coupling and in-plane magnetic field. Indeed, at $k_{R}=0$, the two Fermi surfaces corresponding to spin-polarized electrons are characterized by distinct Fermi momenta $k_{F}^{\pm}$ [see the inset in Fig.\ \ref{fig:2}(b)], which give rise to two spin-dependent \textit{slow} escape rates $\Gamma_{-}^{\pm} = \Gamma \left [1-J_{0} \left (ak_{F}^{\pm}\right )\right ]\approx  \Gamma \left (ak_{F}^{\pm}\right )^{2}/4\ll \Gamma$, resulting in spin splitting of the narrow conduction peak [see Fig.\ \ref{fig:2}(b)]. With SO coupling turned on, the Fermi surfaces are no longer circles with constant radii $k_{F}^{\pm}$, but instead represent closed curves in the $\textbf{k}$ plane, with $k_{F}^{\pm}(\varphi)$ varying strongly along a Fermi surface [see the inset in Fig.\ \ref{fig:5}(b)]. Since all electrons at the Fermi level participate in tunneling, this leads to washing out of spin-resolved features in the conductance lineshape. 

Finally, consider now the role of  Aharonov-Bohm flux through the area enclosed by electron tunneling pathways between left and right 2DEGs at finite angle $\phi$ (see Fig.\ \ref{fig:1}). In Fig.\ \ref{fig:6} we show the conductance evolution as $\phi$ changes  between $\phi=0$ (no flux) to $\phi=\pi/2$ (maximal flux) in both the absence and presence of SO coupling. In either case, the Aharonov-Bohm phase suppresses the interference that causes the narrow peak of the conductance and hence destroys its spin structure. The adverse effect of the Aharonov-Bohm phase on spin-resolved Dicke tunneling is consistent with the spin-independent case \cite{shahbazyan-prb94}.

\section{Conclusions}
\label{sec:5}

In summary, we have considered spin-dependent coherent transport in a double quantum dot system sandwiched between two-dimensional electron gases in doped quantum wells. We have found that for relatively small interdot separation the narrow Dicke conductance peak develops a well resolved spin structures even for very small Zeeman splitting of quantum dot energy levels. We also show that this spin structure is inhibited by SO coupling in a two-dimensional electron gas as well as by Aharonov-Bohm flux through the area enclosed by electron tunneling pathways.


\acknowledgments
This work was supported by the National Science Foundation under Grant No. DMR-1206975 and the CREST center. L.S.P. acknowledges support from the State Committee of Science, Republic of Armenia.


\begin{thebibliography}{}

\bibitem{yacoby-prl95} A. Yacoby, M. Heiblum, D. Mahalu, and H. Shtrikman, Phys. Rev. Lett. \textbf{74}, 4047 (1995).

\bibitem{shuster-nature97} R. Schuster, E. Buks, M. Heiblum, D. Mahalu, V. Umansky, and H. Shtrikman, 
Nature (London) \textbf{385}, 417 (1997).

\bibitem{buks-nature98} E. Buks, R. Schuster, M. Heiblum, D. Mahalu, and V. Umansky, 
Nature (London)  \textbf{391}, 871 (1998).

\bibitem{shahbazyan-prb94}T. V. Shahbazyan and M. E. Raikh, Phys. Rev. B 49, 17123 (1994).

\bibitem{blick-prl01} A. W. Holleitner, C. R. Decker, H. Qin, K. Eberl, and R. H. Blick, 
Phys. Rev. Lett. \textbf{87}, 256802 (2001).

\bibitem{ensslin-prl06} M. Sigrist, T. Ihn, K. Ensslin, D. Loss, M. Reinwald, and W. Wegscheider, 
Phys. Rev. Lett. \textbf{96}, 036804 (2006).

\bibitem{hatano-prl11} T. Hatano, T. Kubo, Y. Tokura, S. Amaha, S. Teraoka, and S. Tarucha, 
Phys. Rev. Lett. \textbf{106}, 076801 (2011). 

\bibitem{loss-prl00} D. Loss and E. V. Sukhorukov, Phys. Rev. Lett. \textbf{84}, 1035 (2000).

\bibitem{gefen-prl01} J. K\"{o}nig and Y. Gefen, Phys. Rev. Lett. \textbf{86}, 3855 (2001).

\bibitem{brandes-pr05} T. Brandes, Phys. Rep. \textbf{408}, 315 (2005), and references therein.


\bibitem{kubala-prb02} B. Kubala and J. K\"{o}nig, Phys. Rev. B \textbf{65}, 245301 (2002).

\bibitem{brandes-prb03}T. Vorrath and T. Brandes, Phys. Rev. B \textbf{68}, 035309 (2003).

\bibitem{orellana-prb03} M. L. Ladron de Guevara, F. Claro, and P. A. Orellana, Phys. Rev. B \textbf{67}, 195335 (2003).


\bibitem{dicke-pr54}R. H. Dicke, Phys. Rev. \textbf{93}, 99 (1954).

\bibitem{devoe-prl96} R. G. DeVoe and  R. G. Brewer, Phys. Rev. Lett. \textbf{76}, 2049 (1996).


\bibitem{shahbazyan-prb98}T. V. Shahbazyan and S. E. Ulloa, Phys. Rev. B \textbf{57}, 6642 (1998).




\bibitem{folk-science03}
J. A. Folk, R. M. Potok, C. M. Marcus, and V. Umansky, Science \textbf{299}, 679 (2003).

\bibitem{hanson-prl03}
R. Hanson, B. Witkamp, L. M. K. Vandersypen, L. H. Willems van Beveren, J. M. Elzerman, and L. P. Kouwenhoven, 
Phys. Rev. Lett. \textbf{91}, 196802 (2003).

\bibitem{hanson-prb04}
R. Hanson, L. M. K. Vandersypen, L. H. Willems van Beveren, J. M. Elzerman, I. T. Vink, and L. P. Kouwenhoven, 
Phys. Rev. B \textbf{70}, 241304(R) (2004).

\bibitem{hanson-prl05}
R. Hanson, L. H. Willems van Beveren, I. T. Vink, J. M. Elzerman, W. J. M. Naber, F. H. L. Koppens, L. P. Kouwenhoven, and L. M. K. Vandersypen, 
Phys. Rev. Lett. \textbf{94}, 196802 (2005).

\bibitem{amasha-prb08}
S. Amasha, K. MacLean, I. P. Radu, D. M. Zumbuhl, M. A. Kastner, M. P. Hanson, and A. C. Gossard,
Phys. Rev. B \textbf{78}, 041306(R) (2008).

\bibitem{otsuka-prb09}
T. Otsuka, E. Abe, Y. Iye, and S. Katsumoto, Phys. Rev. B \textbf{79}, 195313 (2009).

\bibitem{stano-prb10}
P. Stano and P. Jacquod, Phys. Rev. B \textbf{82}, 125309 (2010).

\bibitem{fujisawa-prb14}
M. Yamagishi, N. Watase, M. Hashisaka, K. Muraki, and T. Fujisawa, 
Phys. Rev. B \textbf{90}, 035306 (2014).








\bibitem{datta-prl02}
T. Koga, J. Nitta, H. Takayanagi, and S. Datta,
Phys. Rev. Lett. \textbf{88}, 126601 (2002).

\bibitem{raichev-prb03}
O. E. Raichev and P. Debray,
Phys. Rev. B \textbf{67}, 155304 (2003).

\bibitem{botha-prb03}
A. E. Botha and M. R. Singh,
Phys. Rev. B \textbf{67}, 195334 (2003).

\bibitem{perel-prb03} 
V. I. Perel, S. A. Tarasenko, I. N. Yassievich, S. D. Ganichev, V. V. Belkov, and W. Prettl,
Phys. Rev. B \textbf{67}, 201304 (2003).

\bibitem{tarasenko-prl04} 
S. A. Tarasenko, V. I. Perel, and I. N. Yassievich, Phys. Rev. Lett. \textbf{93}, 056601 (2004).

\bibitem{yu-jap05}
L. Yu and O. Voskoboynikov, J. Appl. Phys. \textbf{98}, 023716 (2005).

\bibitem{glazov-prb05}
M. M. Glazov, P. S. Alekseev, M. A. Odnoblyudov, V. M. Chistyakov, S. A. Tarasenko, and I. N. Yassievich,
Phys. Rev. B \textbf{71}, 155313 (2005).

\bibitem{li-prb06}
W. Li and Y. Guo,
Phys. Rev. B \textbf{73}, 205311 (2006).

\bibitem{raikh-prb06}
V. A. Zyuzin, E. G. Mishchenko, and M. E. Raikh,
Phys. Rev. B \textbf{74}, 205322 (2006).


\bibitem{ye-prb07}
C.-Z. Ye, C.-X. Zhang, Y.-H. Nie, and J.-Q. Liang,
Phys. Rev. B \textbf{76}, 035345 (2007).


\bibitem{rozhansky-prb08}
I. V. Rozhansky and N. S. Averkiev,
Phys. Rev. B \textbf{77}, 115309 (2008).


\bibitem{smoliner-prb12} J. Silvano de Sousa and J. Smoliner,
Phys. Rev. B \textbf{85}, 085303 (2012).


\bibitem{apalkov-jpcm08}
H.-Y. Chen, V. Apalkov, and T. Chakraborty, J. Phys.: Condens. Matter \textbf{20}, 135221 (2008).

\bibitem{yokoyama-prb12} T. Yokoyama and M. Eto,
Phys. Rev. B \textbf{86}, 205305 (2012).





\end{thebibliography}
\end{document}